\documentclass[
superscriptaddress,
final,
reprint,
showkeys,
pre,
aps,
floatfix,
]{revtex4-2}

\usepackage[dvips]{graphicx}
\usepackage[utf8]{inputenc}
\usepackage{amsmath,amssymb,booktabs,latexsym,MnSymbol,bm}
\usepackage{setspace}

\usepackage[%
  colorlinks=true,
  urlcolor=blue,
  linkcolor=blue,
  citecolor=blue
]{hyperref}

\begin{document}

\title{Deep Learning Criminal Networks}

\author{Haroldo V. Ribeiro} 
\email{hvr@dfi.uem.br}
\affiliation{Departamento de F\'isica, Universidade Estadual de Maring\'a -- Maring\'a, PR 87020-900, Brazil}

\author{Diego D. Lopes} 
\affiliation{Departamento de F\'isica, Universidade Estadual de Maring\'a -- Maring\'a, PR 87020-900, Brazil}

\author{Arthur A. B. Pessa}
\affiliation{Departamento de F\'isica, Universidade Estadual de Maring\'a -- Maring\'a, PR 87020-900, Brazil}

\author{Alvaro~F.~Martins} 
\affiliation{Departamento de F\'isica, Universidade Estadual de Maring\'a -- Maring\'a, PR 87020-900, Brazil}

\author{Bruno~R.~da~Cunha} 
\affiliation{Global Investigations, TRM Labs -- San Francisco, CA 94104, United States}
\affiliation{National Police Academy, Brazilian Federal Police -- Brasília, DF 71559-900, Brazil}

\author{Sebasti\'an Gon\c{c}alves} 
\affiliation{Instituto de F\'isica, Universidade Federal do Rio Grande do Sul -- Porto Alegre, RS 91501-970, Brazil}

\author{Ervin K. Lenzi} 
\affiliation{Departamento de F\'isica, Universidade Estadual de Ponta Grossa -- Ponta Grossa, PR 84030-900, Brazil}

\author{Quentin S. Hanley} 
\affiliation{GH and Q Services Ltd, West Studios, Sheffield Road, Chesterfield S41 7LL, United Kingdom}

\author{Matja{\v z} Perc} 
\email{matjaz.perc@gmail.com}
\affiliation{Faculty of Natural Sciences and Mathematics, University of Maribor, Koro{\v s}ka cesta 160, 2000 Maribor, Slovenia}
\affiliation{Department of Medical Research, China Medical University Hospital, China Medical University, Taichung 404332, Taiwan}
\affiliation{Alma Mater Europaea, Slovenska ulica 17, 2000 Maribor, Slovenia}
\affiliation{Department of Physics, Kyung Hee University, 26 Kyungheedae-ro, Dongdaemun-gu, Seoul, Republic of Korea}
\affiliation{Complexity Science Hub Vienna, Josefst{\"a}dterstra{\ss}e 39, 1080 Vienna, Austria}

\begin{abstract}
Recent advances in deep learning methods have enabled researchers to develop and apply algorithms for the analysis and modeling of complex networks. These advances have sparked a surge of interest at the interface between network science and machine learning. Despite this, the use of machine learning methods to investigate criminal networks remains surprisingly scarce. Here, we explore the potential of graph convolutional networks to learn patterns among networked criminals and to predict various properties of criminal networks. Using empirical data from political corruption, criminal police intelligence, and criminal financial networks, we develop a series of deep learning models based on the GraphSAGE framework that are able to recover missing criminal partnerships, distinguish among types of associations, predict the amount of money exchanged among criminal agents, and even anticipate partnerships and recidivism of criminals during the growth dynamics of corruption networks, all with impressive accuracy. Our deep learning models significantly outperform previous shallow learning approaches and produce high-quality embeddings for node and edge properties. Moreover, these models inherit all the advantages of the GraphSAGE framework, including the generalization to unseen nodes and scaling up to large graph structures.
\end{abstract}
\keywords{organized crime, complexity, crime prediction, GraphSAGE, graph convolutional networks}

\maketitle

\section*{Introduction}

Machine learning methods have become increasingly prevalent in scientific investigations across a broad range of disciplines, including materials science~\cite{wei2019machine, butler2018machine}, chemistry~\cite{artrith2021best}, physics~\cite{carleo2019machine}, biology~\cite{tarca2007machine}, and sociology~\cite{molina2019machine}. The recent proliferation of these techniques is tightly related to the rapid growth in the amount of detailed information about diverse systems, as well as the rapid development of artificial intelligence approaches capable of handling the most varied types of data. There is, in fact, a strong tendency in many scientific disciplines towards becoming more and more dependent on methods that are capable of extracting useful knowledge from large-scale and often heterogeneous datasets. Graphs and complex networks are prime examples of such data, and they have recently attracted the attention of researchers as one of the most compelling machine learning paradigms~\cite{zhou2020graph, wu2020comprehensive}. Unlike time series or images, which are arranged in arrays or grid structures, networks have more complex data structures, with nodes representing entities and edges indicating relationships among them without any spatial association. This difference is crucial because machine learning approaches that rely on spatial or temporal relationships, such as convolutional~\cite{goodfellow2016deep} or recurrent neural networks~\cite{abiodun2018state}, are unsuitable for networks. 

The primary challenge for the application of machine learning to network data is thus to encode graph elements, such as edges and nodes, into vector representations -- a process known as graph representation learning~\cite{hamilton2017representation, hamilton2022graph}. Early approaches to extract features from graphs focused on carefully-chosen combinations of network statistics, such as centrality and clustering measures, but they soon proved limited due to their lack of generalization. More recent approaches are, however, much more flexible and can be grouped into two categories~\cite{zhou2020graph}: traditional graph embedding methods and graph neural networks. The first category includes representative embedding algorithms such as DeepWalk~\cite{perozzi2014deepwalk} and node2vec~\cite{grover2016node2vec}, which use random walks over the network to optimize the embedding vectors and make nodes that tend to co-occur in these random walks close in the embedding space. Graph neural networks~\cite{wu2020comprehensive}, in turn, are perhaps the most recent innovation in learning representations of graphs. These deep learning models can generate custom representations based on the message-passing framework (or the graph convolution operator), in which nodes iteratively aggregate information from their local neighborhood to output predictions in machine learning tasks in an end-to-end fashion.

Despite the widespread use of, and the many recent advances in machine learning methods for graphs, applications involving criminal networks are surprisingly scarce. This can likely be attributed to the fact that the complex network framework has only recently entered the toolbox of researchers working with crime data~\cite{d2015statistical, jusup2022social}, although it is already considered an ideal approach to investigating and understanding the intricate associations among criminals~\cite{d2015statistical, luna2020corruption, kertesz2021complexity, granados2021corruption, da2021criminofisica}. Indeed, recent research has demonstrated that patterns exhibited by complex networks related to criminal activities can tie criminal associations not only with individual skills but also with the global structure of these networks~\cite{duijn2014relative, calderoni2017communities, da2018topology, ribeiro2018dynamical, colliri2019analyzing, wachs2019network, da2020assessing, garcia2020ai, solimine2020political, wachs2021corruption, nicolas2021conspiracy, joseph2021ties, martins2022universality}. Similar to evidence at a crime scene, patterns among networked criminals may serve as predictive features for identifying missing links or properties of criminal associations, and they may even provide indications for future criminal behavior. 

However, whether this idea is productive or not remains little explored, with one of the few exceptions being our recent work~\cite{lopes2022machine}. In that work, using data from political corruption, criminal police intelligence, and criminal financial networks, we demonstrated that vector representations of nodes and edges obtained from node2vec combined with shallow statistical learning algorithms (logistic regression and $k$-nearest neighbor) are effective in a series of predictive tasks related to recovering missing criminal partnerships, distinguishing types of associations, predicting the amount of money exchanged among criminal agents, as well as anticipating partnerships during the growth of criminal networks. Building on these same datasets, here, we further improve the accuracy of predictive tasks on criminal networks using deep learning models based on graph convolutional networks. Our results show that these models produce much better embeddings, which in turn yield significantly higher accuracies in most predictive tasks. In particular, we find graph neural networks to improve accuracy by approximately 30\% in a task related to distinguishing among criminal, non-criminal, and mixed relationships in a criminal intelligence network (from 74\% to 99\%), and by approximately 20\% in the accuracy of predicting future partnerships in political corruption networks (from 75\% to 90\%). These deep learning models also improve the coefficient of determination of the relation between the actual and predicted amount of money exchanged among agents in a criminal financial network by approximately 40\% (from 0.64 to 0.90). Moreover, we further demonstrate that graph neural networks can be used to predict whether first-time offenders will become recidivist criminals with an average accuracy of approximately 80\%.

In what follows, we detail these results by first presenting our datasets of criminal networks. We then describe our approach to frame each machine learning problem, including the particular architecture we employ for the graph convolutional networks, and report on the performance of our models in comparison to previous results. Finally, we conclude our article by offering a brief outlook on our findings and some concluding remarks about the promising potential of deep learning methods in the context of criminal networks.

\section*{Datasets}

The empirical data used in our investigation is the same as that reported in Ref.~\cite{lopes2022machine} and refer to four criminal networks. The first two datasets of networks consist of individuals implicated in political corruption scandals that occurred in Spain~\cite{martins2022universality} and Brazil~\cite{ribeiro2018dynamical}. The Spanish corruption network encompasses 437 thoroughly documented cases of corruption that took place between 1989 and 2018, involving 2,695 agents. Similarly, the Brazilian corruption network comprises 65 well-documented corruption scandals that took place between 1987 and 2014, involving 404 agents. In both corruption networks, each node represents an agent engaged in a political scandal, while edges connect pairs of agents who were involved at least once in the same corruption case. In addition to the final stages of these corruption networks (when considering all scandals regardless of when they occurred), we recreate each yearly stage of the growth process of these networks produced by the discovery of scandals over the years. The third dataset is a criminal intelligence network maintained by the Brazilian Federal Police~\cite{da2018topology}. This network is created by using records of criminal investigations conducted by the police such that each of the 23,666 individuals in this network is either a criminal or is suspected of illegal activities related to federal crimes (drugs and arms trafficking, organized bank robbery, environmental crimes, crimes against elections and financial systems, money laundering, among others). The connections among these individuals indicate that they were involved in the same police investigation or shared some personal relationships uncovered during the investigations. Additionally, for the 8,894 individuals belonging to the giant component of this network, the connections among them are classified into three types: criminal (individuals solely related for unlawful purposes), non-criminal (individuals associated for legal purposes such as family or friendship ties), and mixed (individuals that are simultaneously related for legal and illegal purposes such as family members engaged in criminal activities). Finally, the fourth criminal network used in our study refers to a money-laundering investigation conducted by the Brazilian Federal Police from 2008 to 2014~\cite{lopes2022machine}. In this network, the 1,126 nodes represent people or companies involved in the money-laundering case, and the connections indicate the total amount of money exchanged among them regardless of the cash flow direction. 

\section*{Results}

We begin by presenting the essential ingredients that are shared across all the deep learning models based on graph convolutional networks we have used in our machine learning tasks. All our models rely on the GraphSAGE framework~\cite{hamilton2017inductive} which, unlike the original graph convolutional network proposal~\cite{kipf2017semi}, is an inductive framework. This means that models trained on a specific graph structure can be directly applied to a graph with a different structure, as in the case of previously unseen nodes. GraphSAGE also has the advantage of not relying on the entire graph structure ({\it i.e.}, the full-graph Laplacian) to define the convolution operation, making this model scalable for large graphs. Specifically, GraphSAGE samples a fixed number of neighbors for each node up to a given number of hops, and then learns a function that aggregates information from this sampled neighborhood and concatenates it with the node's own information. If $h_u^k$ is the information vector of node $u$ after iteration $k$ with $h_u^0$ being the input node features, we can formulate the GraphSAGE iterative process as
\begin{equation}\label{eq:graphsage}
\begin{split}
    h_{\mathcal{N}_u}^k &= \text{AGG}_{k}\left(\{h_v^{k-1}, \forall v \in \mathcal{N}_u\}\right)\,,\\
    h_{u}^k &= \sigma\left(W^k \cdot \text{CONCAT}(h_{u}^k, h_{\mathcal{N}_u}^k)\right)\,.
\end{split}
\end{equation}
In these equations, $\mathcal{N}_u$ represents the sampled neighborhood from node $u$, $\text{AGG}_{k}$ is the function used to aggregate the information vectors sampled from $u$'s neighborhood into the vector $h_{\mathcal{N}_u}^k$, $\sigma$ is the rectified linear unit (ReLU) activation function, $\text{CONCAT}$ indicates the concatenation operation, and $W^k$ is a weight matrix with learnable parameters. Equation~\ref{eq:graphsage} is iterated from $k=1$ to $k=K$, with $K$ representing the search depth or the maximum number of hops. Each iteration $k$ is often referred to as a convolutional layer of the model and one can view the aggregated information $h_u^k$ as neurons of a fully connected neural network. We use the mean operator as the aggregation function in all our models due to its simplicity and performance in preliminary tests. It is also worth mentioning that this choice is nearly equivalent to the original graph convolutional network~\cite{kipf2017semi}. However, in addition to the inductive advantage, the concatenation operation (which is not present in the original graph convolutional network) represents a kind of skip connection between the GraphSAGE convolutional layers. This feature is now common in several neural network architectures, and usually facilitates the training process and improves the model's generalization and accuracy~\cite{he2016identity}.

\begin{figure*}[!ht]
  \centering
  \includegraphics[width=.8\textwidth, keepaspectratio]{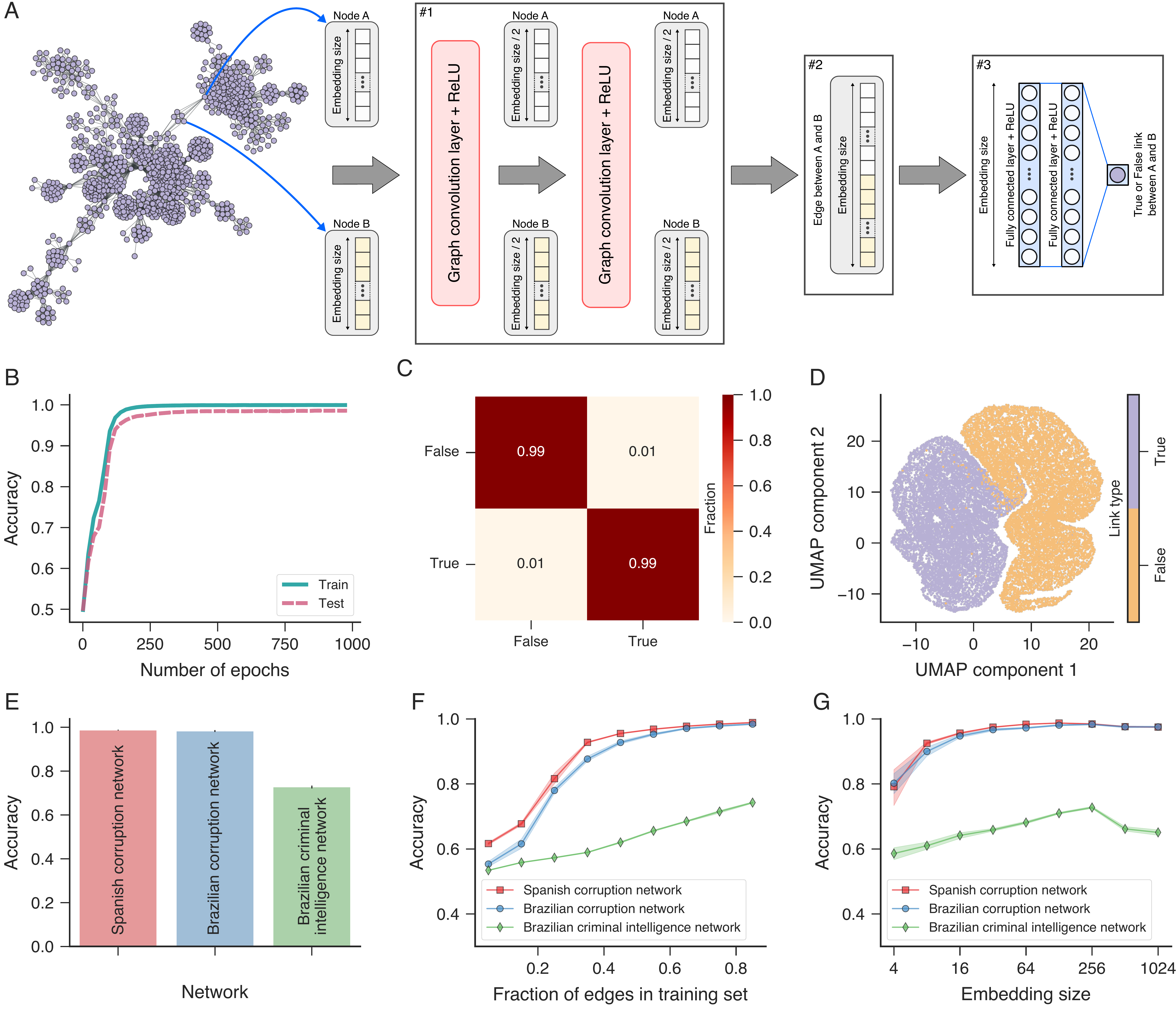}
  \caption{Recovering missing criminal partnerships with graph neural networks. (A) Schematic representation of the neural network architecture used for recovering a fraction of edges randomly removed from criminal networks. Every node has a feature vector of a given size (the embedding size) that passes through a sequence of two graph convolution (GraphSAGE with mean aggregator function) layers combined with rectified linear unit (ReLU) activation functions. The output of these convolution layers is a vector with half the size of each node's feature vector. These vectors are stacked together to represent a possible link between a pair of nodes (A and B in this example). This vector representation for a possible edge is passed through two fully connected layers with the number of nodes equal to the embedding size. Finally, the edge classification (whether a true or a false edge between nodes A and B) takes place in the output layer via a sigmoid activation function. (B) Example of training and testing scores (fraction of correct classifications) as a function of the number of epochs used during the training stage. (C) Example of confusion matrix obtained when applying the trained model to the test set (rows indicate actual labels). (D) Visualization of typical edge embeddings generated by our model where the different colors indicate true (purple) and false (orange) connections. We have considered the output of the second layer of the fully connected layers as the final embedding for a possible edge and used the uniform manifold approximation and projection (UMAP) technique to represent these vectors in a two-dimensional space. Results in panels (B-D) refer to one realization of the training procedure using the final stage of the Spanish corruption network and an embedding size with $256$ dimensions. The training set comprises $80$\% of the true network edges (randomly sampled) and the same number of false links randomly generated, while the test set (which is never used during the training stage) is composed of the $20$\% remainder of the true network edges and the same number of false links randomly generated. (E) Average model accuracy over the test set (with $20$\% of edges) calculated from twenty independent realizations of the training process (embedding size with $256$ dimensions) for the Spanish corruption, Brazilian corruption, and Brazilian criminal intelligence networks (the tiny error bars represent $95$\% confidence intervals). (F) Average model accuracy over the test sets as a function of the fraction of edges in the training sets for each criminal network and embedding size with $256$ dimensions. (G) Average model accuracy over the test sets (with $20$\% of edges) as a function of the embedding size for each criminal network. In the two previous panels, markers represent the average accuracy estimated from twenty independent realizations of the training process while the shaded regions stand for $95$\% confidence intervals.
  }
  \label{fig:1}
\end{figure*}

After introducing the fundamental concepts of the GraphSAGE framework, our focus now turns to the task of recovering missing partnerships in criminal networks. Specifically, we consider the final stages of the Spanish and Brazilian corruption networks and the criminal intelligence network. To create our training sets, we randomly sample a fraction of the actual edges ($f_{\text{train}}$) from these networks and generate an equivalent number of false connections. The remaining edges ($f_{\text{test}} =1-f_{\text{train}}$) are combined with a sample of randomly generated false connections of the same size to form the testing sets. Our objective is thus to develop a model capable of accurately identifying true and false connections within these testing sets. Figure~\ref{fig:1}A illustrates the architecture of our model. It begins by subjecting the feature vectors of two nodes (A and B) to a two-layer GraphSAGE convolutional neural network. Next, we concatenate the information vectors resulting from the graph convolutions and feed them through two fully connected layers with ReLU activation functions. The first graph convolution layer reduces the dimension of input features to half its initial size, such that the concatenated information from the two nodes has the same dimension as the input features. Finally, we feed the information from the last fully connected layer into a single-neuron layer with a sigmoid activation function (corresponding to logistic regression), which outputs the final classification -- a prediction for whether nodes A and B are connected or not. We initially use 80\% of network edges for the training sets, and to generate the input node features ($h_u^0$), we consider the embeddings produced by node2vec~\cite{grover2016node2vec} with 256 dimensions and other parameters fixed to the default settings (walk length equals 5, number of walks per node equals 10, and bias parameters set to 1) for the criminal networks recreated using only edges in the training sets. 

We use the Adam stochastic gradient descent method~\cite{kingma2014adam} with a learning rate of 0.001 and employ the binary cross-entropy as the loss function to optimize the model parameters. To mitigate the risk of overfitting, we incorporate an early stopping regularization procedure with a patience level of $100$ epochs and include an L2 regularization term in the loss function with a hyperparameter of $0.001$. Figure~\ref{fig:1}B shows the accuracy (fraction of correctly classified edges) as a function of the number of training epochs calculated from the training and testing sets of the Spanish corruption network. The training score reaches saturation around the maximum value after roughly 500 epochs, while the testing score is slightly lower. The confusion matrix presented in Figure~\ref{fig:1}C further corroborates the efficacy of our model and demonstrates its ability to accurately distinguish false and true connections within the testing set. This robust discrimination power indicates that our graph convolutional network produces embeddings for false and true connections that are distinctly separated. To visualize these edge embeddings, we consider the layer's output just before the classification layer as edge embeddings and project these 256-dimensional vectors into two dimensions using the uniform manifold approximation and projection (UMAP) method~\cite{mcinnes2018umap, mcinnes2018umapsoftware}. Figure~\ref{fig:1}D displays the projected embeddings when considering the edges in the testing set of the Spanish corruption network. True and false edges occupy distinct regions of the UMAP plane with minimal overlapping, thereby explaining the model's excellent performance and the expressive capabilities of its simple architecture.

We apply the same model used for the Spanish corruption network to the Brazilian corruption and criminal intelligence networks and obtain comparable good results. To summarize these findings, we calculate the average accuracy in distinguishing false and true connections over twenty random instances of the train-test split procedure. Figure~\ref{fig:1}E presents these average scores for the three criminal networks. Our models exhibit nearly perfect classification performance for both corruption networks ($99$\% for the Spanish and $98$\% for the Brazilian), while the accuracy for the criminal intelligence network is $73$\%. We also examine how the average scores depend on the fraction of edges in the training sets. We vary the fraction of edges used to train the models from 5\% to 85\% (in increments of 10\%) and calculate the average accuracy within the testing sets over twenty random instances of the train-test split procedure. As shown in Figure~\ref{fig:1}F, the accuracy of our model improves with an increase in the number of training samples for the three criminal networks, but with a distinct pattern among the networks. The scores for corruption networks approach saturation after considering about 60\% of the data, while the accuracy for the criminal intelligence network monotonically increases with the fraction of edges in the training set. Furthermore, we verify how the dimension of the input node features produced by node2vec affects the performance of our models. To do so, we consider the 80\%-20\% train-test split and vary the dimension of the input node features from 4 to 1024 with log-spaced increments ($2^i~\forall i \in 2,3,\dots,10$). We train the models for each of these dimensions and calculate their average accuracies over twenty random instances of the train-test split procedure, as shown in Figure~\ref{fig:1}G. Again, we find distinct behaviors for corruption and criminal intelligence networks. While the scores of corruption networks rapidly approach maximum values for embedding sizes around 256 and slightly decrease for larger embedding sizes, the accuracy for the criminal intelligence network approximately linear increases up to an embedding size equal to 256 and significantly decreases for larger dimensions. These differences, combined with the smaller accuracy obtained for the intelligence network, indicate that learning good embeddings for this network is significantly challenging compared to the case of corruption networks. Moreover, comparable to our previous work~\cite{lopes2022machine}, the accuracies observed for the corruption networks are slightly higher (99\% vs. 98\% and 98\% vs. 96\% for the Spanish and Brazilian corruption networks, respectively), while performance obtained with the criminal intelligence network is significantly worst ($73$\% vs. $87$\%). We also note that the learning curves of our deep-learning models (Figure~\ref{fig:1}F) are less steep than the ones obtained from our previous shallow-learning approach. We believe this result reflects the fact that deep learning models tend to be much more data-intensive to train. Additionally, we remark that the criminal intelligence network exhibits more complex and less redundant structures compared with corruption networks that are formed by complete graphs connected by the recidivism of agents~\cite{martins2022universality}. This leads us to conjecture that exposing our model to larger training samples related to the criminal intelligence network, if they were available, would allow it to further improve its performance. 

\begin{figure*}[!ht]
  \centering
  \includegraphics[width=0.8\textwidth, keepaspectratio]{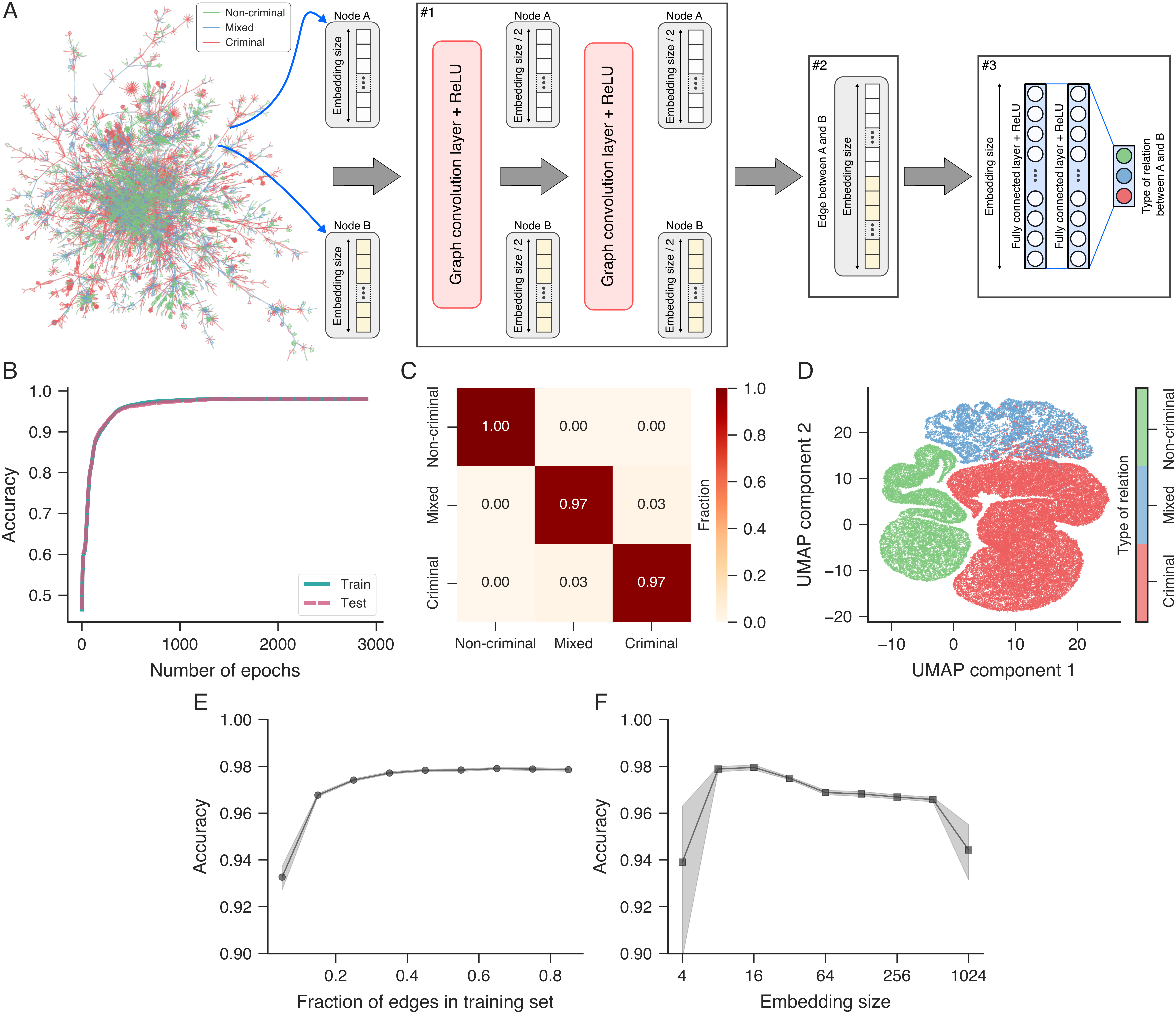}
  \caption{Predicting the type of relationship among agents in criminal networks with graph neural networks. (A) Schematic representation of the neural network architecture used for determining the type of relationship among agents in the Brazilian criminal intelligence network. This architecture is similar to the one used for recovering missing links. The only difference is in the last layer, which is now composed of three nodes with softmax activation functions. These output nodes represent the three possible types of relationships: criminal, mixed, and non-criminal. (B) Example of training and testing scores (fraction of correct classifications) as a function of the number of training epochs. (C) Example of confusion matrix obtained when applying the trained model to the test set of network edges (rows indicate actual labels). (D) Visualization of typical edge embeddings generated by our model where the different colors indicate the three types of relationships (red: criminal, blue: mixed, and green: non-criminal). This visualization is obtained by considering the output of the second layer of the fully connected layers as the embedding for edges in the test set and by mapping these vectors into a two-dimensional space with the uniform manifold approximation and projection (UMAP) technique. Results in panels (B-D) are from one realization of the training procedure, initial embedding size with $16$ dimensions, and a training set comprising $80$\% of the network edges (randomly sampled and stratified by the three classes). The $20$\% remainder of the network edges are used as the test set. (E) Average model accuracy over the test set as a function of the fraction of edges in training sets and embedding size with $16$ dimensions. (G) Average model accuracy over the test set (with $20$\% of edges) as a function of the embedding size. In the two previous panels, markers represent the average accuracy estimated from twenty independent realizations of the training process while the shaded regions stand for $95$\% confidence intervals.
  }
  \label{fig:2}
\end{figure*}

Moving to our second application, we focus on the giant component of the criminal intelligence network, for which we have information about the nature of associations among agents. Our goal is now to differentiate between criminal, mixed, and non-criminal relationships. Figure~\ref{fig:2}A displays the architecture of the model proposed for this task. This model differs from the previous one only in the output layer, which consists of three neurons, one for each type of association, with softmax activation functions. For one instance of the 80\%-20\% train-test split, we optimize the model parameters using the same strategy as the previous task with the categorical cross-entropy as the loss function, and input features generated from node2vec with 16 dimensions. The training and testing scores as a function of the number of training epochs are presented in Figure~\ref{fig:2}B. We observe that these scores saturate after about 2000 epochs at approximately 98\% accuracy. Despite the imbalanced edge classes (54\% criminal, 22\% mixed, and 24\% non-criminal), we do not use any strategy to balance the class distribution in the training set. Still, as shown by the confusion matrix of Figure~\ref{fig:2}C, our model is nearly equally effective in distinguishing among the three types of relationships. In this particular instance of the train-test split, the model has only misclassified 3\% of criminal and mixed relationships. Compared to our previous shallow learning approach based on the $k$-nearest neighbors classifer~\cite{lopes2022machine}, which yielded an overall accuracy of approximately 74\%, the current deep learning model represents a significant improvement. This improvement is particularly impressive for discriminating mixed relationships, for which the previous approach exhibited an accuracy of only 55\%. Once again, this high performance can be directly attributed to the quality of edge embeddings produced by the model. To visualize these embeddings in the test set, we consider the output of the last layer before the classification layer and use the UMAP to project these 16-dimensional vectors into a plane, as shown in Figure~\ref{fig:2}D. These results confirm the high quality of the embedding produced and show that only a small portion of mixed and criminal edges overlap in this two-dimensional projection.

We further verify how the performance of our model changes with the number of training samples. We vary the fraction of edges in the training sets from 5\% to 85\% in increments of 10\%, and calculate the average accuracy within the testing sets over twenty random instances of the train-test split procedure. The learning curve of our model is depicted in Figure~\ref{fig:2}E, where we observe that the accuracy saturates after considering about half of the network edges in the training sets. This behavior also differs from what was observed in our previous shallow learning approach~\cite{lopes2022machine}, which exhibited a monotonically increasing accuracy with the fraction of edges in the training sets. Additionally, we investigate the role of the embedding size used with the node2vec algorithm for producing the input node features. Specifically, we consider the 80\%-20\% train-test split again, vary the embedding size of node2vec from 4 to 1024 with log-spaced increments ($2^i~\forall i \in 2,3,\dots,10$), and calculate the average accuracy in the testing sets over twenty random instances of the train-test split procedure for each embedding size. These scores are presented in Figure~\ref{fig:2}F, where even small embedding sizes yield high accuracies. However, the maximum accuracy occurs when generating input node features with 16 dimensions.

\begin{figure*}[!ht]
  \centering
  \includegraphics[width=0.8\textwidth, keepaspectratio]{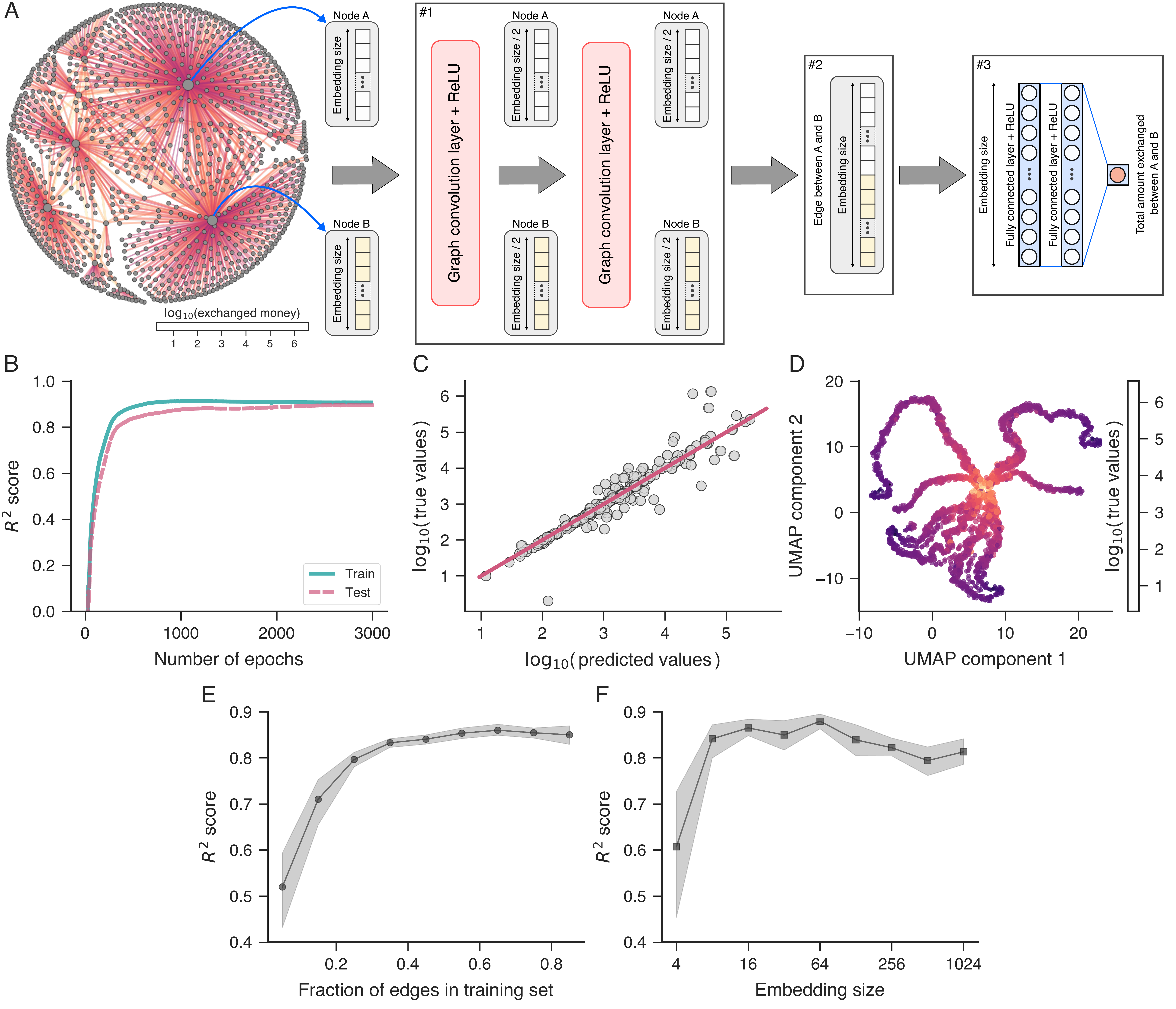}
  \caption{Predicting the amount of money exchanged among agents in a criminal financial network with graph neural networks. (A) Schematic representation of the neural network architecture used for predicting the total amount of money exchanged between pairs of agents in a criminal financial network. This architecture is similar to the ones used for recovering missing links and predicting the type of relationship. The only difference is in the last layer, where a linear activation function is used for outputting the expected amount of money exchanged between a pair of nodes. (B) Example of training and testing scores (coefficient of determination $R^2$ between the actual and predicted values on base-10 logarithmic scale) as a function of the number of epochs used during the training stage. (C) Example of relationship between the true and predicted base-10 logarithm values of the amount of money exchanged among agents in the test set of the criminal financial network (the continuous line is the 1:1 relationship). (D) Visualization of typical edge embeddings generated by our model where the color code indicates the actual values for the amounts of money. This visualization is obtained by considering the output of the second layer of the fully connected layers as the embedding for edges in the test set and by mapping these vectors into a two-dimensional space with the uniform manifold approximation and projection (UMAP) technique. Results in panels (B-D) are from one realization of the training procedure, initial embedding size with $32$ dimensions, and a training set comprising $80$\% of the network edges (randomly sampled). The $20$\% remainder of the network edges are used as the test set. (E) Average model $R^2$ score over the test set as a function of the fraction of edges in training sets and embedding size with $32$ dimensions. (G) Average model $R^2$ score over the test set (with $20$\% of edges) as a function of the embedding size. In the two previous panels, markers represent the average $R^2$ score estimated from twenty independent realizations of the training process while the shaded regions stand for $95$\% confidence intervals.
  }
  \label{fig:3}
\end{figure*}

For our third task, we examine the criminal financial network associated with a money-laundering investigation. Our objective is to predict the logarithm of the total money exchanged among agents in this network. To accomplish this, we employ the same architecture as in our previous models but modify the output layer to consist of a single neuron with a linear activation function, representing linear regression. The architecture of our model is depicted in Figure~\ref{fig:3}A. Moreover, we follow the same optimization procedure as before and first consider one instance of the 80\%-20\% train-test split. The input features are generated using node2vec with 32 dimensions, and we train our model using the mean squared error (also known as the squared L2 norm) as the loss function. Figure~\ref{fig:3}B shows the coefficient of determination ($R^2$ score) of the association between true and predicted values in the training and testing sets as a function of the number of training epochs, which saturates at approximately 0.85 after around 2000 training epochs. Figure~\ref{fig:3}C displays the relationship between the true and predicted values estimated for the testing set using the trained model. Notably, the model's performance is significantly superior to that obtained from our previous shallow learning approach based on a $k$-nearest neighbors classifier~\cite{lopes2022machine}, which yielded an $R^2$ score of approximately 0.64. We also apply the UMAP algorithm to the output of the last layer before the regression layer to visualize the embeddings generated by the model. Figure~\ref{fig:3}D shows this visualization, where we use a color map that refers to the logarithm of the total amount of money of the corresponding edges to color each data point. We observe that as we move radially away from the center of the UMAP projection, the values associated with the edges tend to decrease, illustrating the high quality of the embeddings produced by our model.

To assess the reliability of the trained model, we investigate the dependence of the $R^2$ score with the fraction of edges used during the training stage. Figure~\ref{fig:3}E shows the learning curve of our model estimated from testing sets with input features obtained from node2vec with 32 dimensions. Each data point represents the average value of the $R^2$ score computed from twenty random instances of the train-test split procedure, with shaded regions indicating 95\% confidence intervals. The results show that the $R^2$ score approaches saturation after including approximately half of the edges in the training sets. Additionally, we investigate the impact of the embedding size used with node2vec for generating input features on the $R^2$ score. Figure~\ref{fig:3}F shows the average value of the $R^2$ score calculated from testing sets over twenty random instances of the 80\%-20\% train-test split as a function of the embedding size. As depicted in this figure, we note that embedding sizes ranging from 8 to 64 dimensions yield similarly accurate predictions for the logarithm of the total money exchanged among agents in the criminal financial network. Embedding sizes below 8 result in considerably lower $R^2$ scores, while embedding sizes above 64 dimensions only produce slightly lower scores.

\begin{figure*}[!ht]
  \centering
  \includegraphics[width=0.8\textwidth, keepaspectratio]{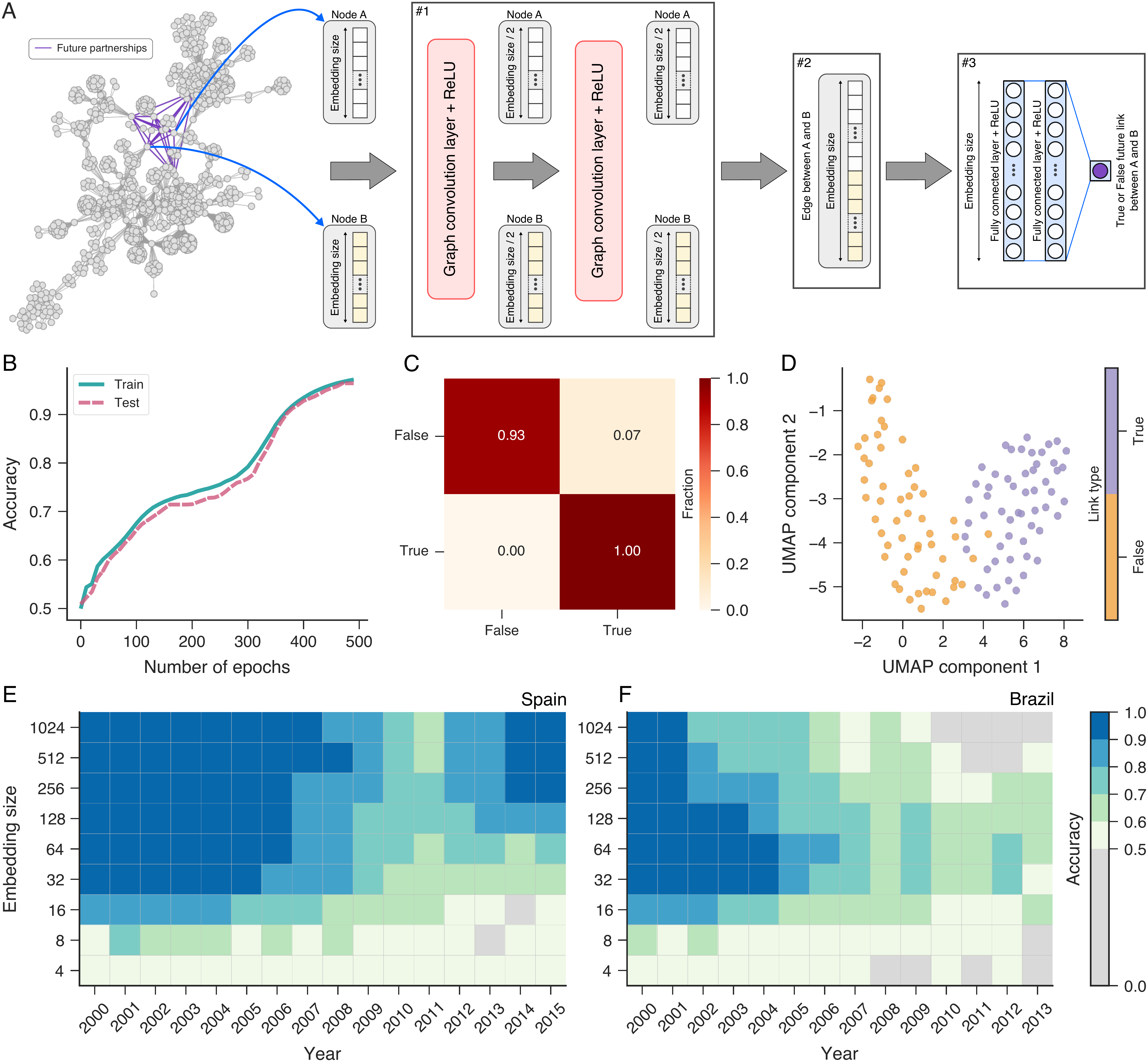}
  \caption{Predicting future partnerships among known agents in corruption networks. (A) Schematic representation of the neural network architecture used to determine future links among agents in corruption networks. This architecture is the same one used for recovering missing links and comprises a block of two graph convolution layers, followed by a concatenation of node embeddings, which are then forwarded into a block of fully connected layers whose output node is responsible for the edge classification. (B) Example of training and testing scores (fraction of correct classifications) as a function of the number of epochs used during the training stage. (C) Example of confusion matrix obtained when applying the trained model to the test set (rows indicate actual labels). (D) Visualization of typical edge embeddings generated by our model where the different colors indicate true (purple) and false (orange) connections. We have considered the second layer of the fully connected layers as the final embedding for a possible edge and used the uniform manifold approximation and projection (UMAP) technique to represent these vectors in a two-dimensional space. Results in panels (B-D) represent one realization of the training procedure using the Spanish corruption network with scandals that occurred up to the year $2014$ and an embedding size with $256$ dimensions. The training set comprises all network edges that occurred up to the year $2014$ and the same number of false links randomly generated. The test set includes edges among known nodes (those emerging before 2014) that will arise in the network's future (edges occurring after 2014) and the same number of randomly generated false links that do not appear in the network's future. (E-F) Average model accuracy over the test sets when considering different years of the Spanish and Brazilian corruption networks as well as different embedding sizes. Each cell in these matrix representations indicates the average accuracy (represented by a color code where gray cells indicate nonsignificant accuracies) over the test set estimated from twenty independent realizations of the training process for a given threshold year (columns) and a given embedding size (rows).
  }
  \label{fig:4}
\end{figure*}

Our fourth task tackles the more challenging problem of predicting future criminal partnerships in corruption networks. We focus on the two corruption networks because for them we have information on their growth over the years, wherein new nodes and edges are added with the discovery of new corruption scandals. Assuming that $G_y$ represents one of these networks after incorporating all scandals up to the year $y$, we consider all network edges and a random sample of the same size of false connections as the training set. Next, we extract all future connections among nodes already present in $G_y$ and a random sample of false future connections of the same size to generate the testing set. Our goal is to train a model capable of distinguishing between true and false future criminal partnerships in corruption networks. To do so, we consider the same model architecture used to recover missing relationships in our first task, as shown in Figure~\ref{fig:4}A. We optimize the model parameters with the same procedures used before and use the binary cross-entropy as the loss function. However, because the set of true and false future connections is smaller than the data used in our previous tasks (hundreds versus thousands of samples), we intensify the regularization procedures by setting the patience level to $10$ epochs and the L2 regularization hyperparameter to $0.002$ in order to avoid overfitting. 

Initially, we consider the Spanish corruption network with all scandals occurring up to the year $y=2014$. We further use node2vec with 256 dimensions to generate the input features of all nodes. Figure~\ref{fig:4}B shows the evolution of the train and test scores (accuracy) with the number of training epochs, where we observe the scores trend towards high accuracy levels. We also estimate the confusion matrix by applying the trained model to the test set. The results of Figure~\ref{fig:4}C show that the model correctly identifies all true future criminal partnerships and misclassifies 7\% of false future criminal partnerships as true ones. This pattern is however not consistent across different realizations of the train-test split procedure nor across different years. Overall, we find that trained models are equally good at discriminating between true and false future connections. Once again the good performance of our model can be attributed to the good quality of edge embeddings it produces. Figure~\ref{fig:4}D visualizes these embeddings obtained from the testing sets after applying the UMAP algorithm to the output of the last layer before the classification layer. Similarly to the previous classification tasks, the embeddings for true and false future connections tend to occupy distinct regions of the UMAP projection.

To systematically investigate the performance of our model over the evolution of the two corruption networks, we consider all network stages between the years 2000 and 2015 of the Spanish network and between 2000 and 2013 for the Brazilian network. For each of these stages, we replicate the processes previously used for the year 2014 of the Spanish network. To produce the input features, we vary the node2vec embedding size from 4 to 1024 in steps logarithmically spaced ($2^i~\forall i \in 2,3,\dots,10$). For each pair of year and embedding size, we generate twenty instances of the random sampling strategy for generating false connections. We then train the model using the same settings employed for the Spanish network in the year 2014 and calculate the average accuracy from testing sets for each embedding size and year of the two corruption networks. These results are depicted in a matrix plot form in Figures~\ref{fig:4}E and \ref{fig:4}F for the Spanish and Brazilian corruption networks, respectively. Each cell corresponds to the average accuracy for a given value of network year (fixed for each column of the matrix) and embedding size (fixed for each line of the matrix). The color code is such that nonsignificant accuracy levels are indicated by gray cells. 
 
We observe that embedding sizes lower than 16 dimensions yield models with accuracy levels close to the threshold of significance, or even nonsignificant in particular years. On the other extreme, very large embedding sizes also tend to produce less than optimal accuracies, especially for the Brazilian corruption network. Intermediate embedding sizes (256 dimensions for the Spanish and 64 for the Brazilian network) yield the highest accuracies. These optimal accuracies are significantly higher than those obtained from our previous shallow learning approach based on logistic regressions~\cite{lopes2022machine}. That approach yielded an average accuracy across the years of approximately 80\% for the Spanish network (versus 90\% of the current approach) and approximately 65\% for the Brazilian network (versus 80\% of the current approach). We further note that the accuracies pass to a minimum between 2009 and 2012 for the Spanish network and between 2005 and 2009 for the Brazilian network. This behavior was also observed with the shallow learning approach~\cite{lopes2022machine} and reflects a percolation-like transition these networks undergo during those time periods~\cite{martins2022universality}. These transitions are characterized by the coalescence of large network components, as well as by the inclusion of several new scandals. For the Spanish network, the accuracy returns to a similarly high level, but the Brazilian network behaves differently and remains at significantly lower accuracy levels after the transition. While it is challenging to precisely formulate an explanation for this difference, these results suggest that partnerships among agents in the Brazilian corruption network are more random compared to the Spanish case.

\begin{figure*}[!ht]
  \centering
  \includegraphics[width=0.8\textwidth, keepaspectratio]{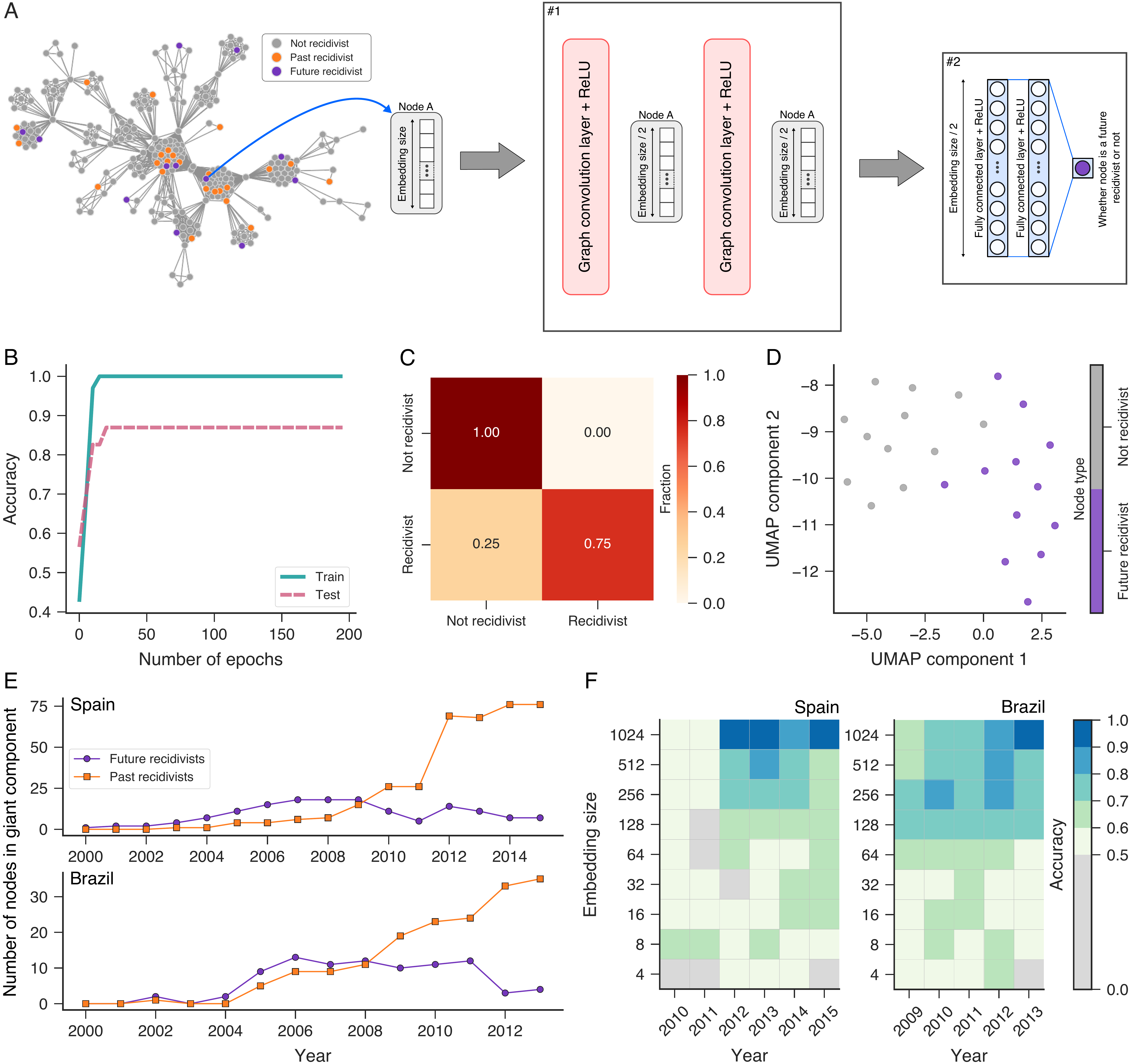}
  \caption{Predicting future recidivist agents in corruption networks. (A) Schematic representation of the neural network architecture used to determine whether an agent will become a recidivist in the network's future (that is, whether a criminal will be involved again in future corruption scandals). This architecture is similar to the ones used in classification tasks at the edge level, but instead of concatenating node representations for creating edge representations,  we directly forward the output of graph convolution layers to the two fully connected layers. The node classification occurs in the output layer via a sigmoid activation function. (B) Example of training and testing scores (fraction of correct classifications) as a function of the number of epochs used during the training stage. (C) Example of confusion matrix obtained when applying the trained model to the test set of network nodes (rows indicate actual labels). (D) Visualization of typical node embeddings generated by our model where the different colors indicate whether nodes are future recidivists (purple) or not (gray). This visualization is obtained by considering the output of the second layer of the fully connected layers as the embedding for nodes in the test set and by mapping these vectors into a two-dimensional space with the uniform manifold approximation and projection (UMAP) technique. Results in panels (B-D) represent one realization of the training procedure using the giant component of the Brazilian corruption network with scandals that occurred up to $2011$ and an embedding size with $256$ dimensions. The training set comprises all nodes that emerged up to the year $2011$ and labels indicating whether or not these nodes are recidivist agents. The test set comprises nodes that will become recidivist agents in the network's future (after $2011$) and the same number of randomly sampled nodes that will not become recidivists. (E) Number of future and past recidivist agents in the giant component of the Spanish and Brazilian corruption networks as a function of time. (F) Average model accuracy over the test sets when considering different years of the Spanish and Brazilian corruption networks as well as different embedding sizes. Each cell in these matrix representations indicates the average accuracy (represented by a color code where gray cells indicate nonsignificant accuracies) over the test set estimated from twenty independent realizations of the training process for a given threshold year (columns) and a given embedding size (rows).
  }
  \label{fig:5}
\end{figure*}

Thus far, we have focused on predictive tasks related to edge features. In a final task, we shift our attention to predicting a node property. Specifically, we address the problem of identifying future recidivist agents in corruption networks, which are agents involved in more than one corruption scandal. Prior research has established that the recidivist rate in a model of corruption networks is a critical parameter for the network structure~\cite{martins2022universality}. Recidivist rates that fall below a given critical value lead to excessively fragmented corruption networks, while rates exceeding this critical value result in overly connected networks~\cite{martins2022universality}. Intriguingly, real corruption networks operate in close proximity to this critical point~\cite{martins2022universality}. In addition to its significance for network structure, the identification of potential recidivist agents can be of great interest to police intelligence operations. To properly formulate this problem and produce the training sets, we consider all past recidivist and non-recidivist agents present in the giant components of the two corruption networks $G_y$, after including all scandals up to a specified year $y$. In turn, our testing sets are generated by considering all agents belonging to the giant component of $G_y$ that will become recidivists in the network's future and a random sample of the same size of agents who will not become recidivists (agents in this sample are excluded from the training sets). In contrast to the previous tasks, where input features from two nodes are simultaneously used, the model architecture we use for identifying recidivist agents only employs the input features for a single node. This architecture is depicted in Figure~\ref{fig:5}A. First, the feature vector of a given node (A) is subjected to a two-layer GraphSAGE convolutional neural network. In this network, the first layer reduces the dimension of the input features to half of its initial size. The outcome of these graph convolutions is then passed through two fully connected layers with ReLU activation functions. Finally, the information from the last fully connected layer is fed into a single-neuron layer with a sigmoid activation function, corresponding to logistic regression. This layer outputs the node classification, predicting whether the node is a recidivist or not.

To illustrate the training process, we consider the giant component of the Brazilian corruption network with scandals up to 2011, which is also illustrated in Figure~\ref{fig:5}A. In this figure, gray nodes represent non-recidivist agents, orange nodes indicate past recidivists, and purple nodes represent agents that will become recidivists in the future of the network. We optimize the model parameters using the binary cross-entropy as the loss function, along with an L2 regularization term with a hyperparameter set to $0.001$. To mitigate the risk of overfitting, we further adopt an early stopping regularization procedure with a patience level of 5 epochs. Moreover, we use node2vec with 256 dimensions to generate input features for all nodes. Figure~\ref{fig:5}B presents the accuracy in the training and testing sets, while Figure~\ref{fig:5}C reports the confusion matrix estimated from the testing set. The accuracy in the training set quickly approaches the maximum value, while the testing score plateaus at a lower level, around 88\%. The confusion matrix shows that the model correctly identifies all non-recidivist agents, but misclassifies one-fourth of recidivist agents as non-recidivists. As in the previous tasks, we consider the output of the last layer before the classification layer as the node embedding and apply the UMAP algorithm to project these vectors into a two-dimensional space. Figure~\ref{fig:5}D displays this projection estimated from the testing set. We observe that non-recidivist and future recidivist agents tend to occupy distinct regions of the UMAP plane. However, the boundary between the two classes is significantly fuzzier than those we have observed in our previous classification tasks. This not-so-clear separation between non-recidivist and future recidivist agents translates into the misclassifications observed in Figure~\ref{fig:5}C.

To extend this analysis to the different years of the Brazilian and Spanish corruption networks, we first calculate the evolution of the number of future and past recidivist agents in their giant components. The results of Figure~\ref{fig:5}E show that having a nonnegligible number of past recidivist agents in both networks takes some time. Specifically, the number of past recidivists surpasses that of future recidivists after 2009 for the Spanish network and after 2008 for the Brazilian network. Therefore, in order to ensure sufficient training samples, we limit our attention to the giant components from the years 2010 and 2009 for the Spanish and Brazilian networks, respectively. For each stage of the networks from these years, we optimize the model parameter for different embedding sizes in the node2vec algorithm ($2^i~\forall i \in 2,3,\dots,10$), which we use to generate the input node features. Furthermore, we replicate the training processes in twenty random instances to calculate the average accuracy of the model in the testing sets. The resulting scores are presented in the matrix plots of Figure~\ref{fig:5}F for both the Spanish and Brazilian networks. We observe that the fractions of correct classifications are almost always lower than 70\% for embedding sizes smaller than 128 dimensions, with some cases not reaching the significance level of 50\%. Only higher embedding sizes produce scores above 80\%. Additionally, we verify that the misclassification of future recidivists as non-recidivist agents reported in Figure~\ref{fig:5}C is not an incidental feature of the Brazilian network in 2011, but instead, it is a pattern we observe in all years of both networks. In other words, our model exhibits a trend of systematically not identifying some future recidivist agents. We believe this failure indicates that the recidivism of criminal agents cannot be solely attributed to the topological properties of corruption networks. In our case, we only have information about the structure of these criminal networks. However, it would be an intriguing possibility for future research to investigate whether additional information, such as the political parties or geographic locations of individuals involved in corruption scandals, would result in higher scores, or if it would reveal that recidivism in criminal networks has an intrinsic non-deterministic nature.

\section*{Discussion}

We have presented a series of applications of graph neural networks for predictive tasks related to various properties of criminal networks. By using data from two corruption networks and one criminal intelligence network, we have demonstrated that a simple architecture consisting of two GraphSAGE convolutional layers, followed by two fully connected layers and a final output layer, can accurately recover missing criminal partnerships in a static setting where some edges are randomly removed. A very similar model has also proven effective in distinguishing among criminal, mixed, and non-criminal associations between individuals in the giant component of a criminal intelligence network. In addition to achieving good results with classification and link prediction tasks, we have shown that our models can be easily adapted to regression tasks that seek to determine the total amount of money exchanged among agents in a criminal financial network. In this case, our models have also produced high-quality predictions. Beyond being useful in static settings, our models based on GraphSAGE convolutional networks also exhibit predictive power for anticipating dynamic properties of edges and nodes during the growth dynamics of corruption networks. We have demonstrated that these models can predict future partnerships among known agents in corruption networks with significant accuracy. Additionally, our models proved useful in identifying future recidivist agents in corruption networks, again with significant accuracy. In all predictive tasks, we have verified that the quality of our predictions is directly attributed to the quality of the node and edge embeddings produced by our models. Specifically, using a dimension reduction technique, we have shown that the vector representations produced by our models for different classes related to edge and node properties and the existence or absence of links between nodes tend to occupy distinct regions of the embedding space. Similarly, in the case of the regression task related to the total amount of money exchanged among agents in a criminal financial network, we have found the embeddings produced by our models to consistently reflect the predicted property.

With the exception of the task of predicting future recidivist agents in corruption networks, for which we have no comparative work to discuss, and missing links in a criminal intelligence network, we have demonstrated that all of our deep learning models significantly outperform our previous shallow learning approach based on node2vec, logistic, and $k$-nearest neighbor models~\cite{lopes2022machine}. Specifically, we have observed an improvement of approximately 30\% in the accuracy of predicting the type of relationships in a criminal intelligence network (from 74\% to 99\%), of about 20\% in the accuracy of anticipating future partnerships in corruption networks (from 80\% to 90\% in the Spanish network and from 65\% to 80\% in the Brazilian network), and of approximately 40\% in the regression task of predicting the amount of money exchanged among agents in a criminal financial network (adjusted $R^2$ from 0.64 to 0.90). Furthermore, our models naturally inherit all the computational advantages of the GraphSAGE framework, including its inductive and expressive capabilities, as well as the ability to scale to large graph structures. The inductive property is particularly interesting for practical applications in scenarios where criminal networks are growing, as it allows the trained model to be directly applied to previously unseen nodes. Although the use of node2vec to generate the input features of each node limits this property in our models, we believe that one can restore this capability by directly aggregating input features in the neighborhood of previously unseen nodes and attributing these aggregated vectors as the input features to unseen nodes. Moreover, these models can be easily adapted to consider simpler topological properties as input features, which can be promptly evaluated for previously unseen nodes, or even more interestingly, to include other node features not directly related to the network. Simple examples include demographic information about criminal agents, such as age, location, and gender, but could also scale to more complex features such as text embeddings of the available criminal files or other textual information related to police investigations.

In conclusion, our work showcases the immense potential of deep learning models in exploring, predicting, and even forecasting properties of criminal networks. The models developed here not only provide significant improvements over previous approaches but may also inspire further research in this area. With the increasing complexity of criminal activities, the applications of such models could aid law enforcement agencies in their investigations, and in doing so provide valuable insights and guidance.

\section*{Data availability}
All data used in this article are available from Refs.~\cite{ribeiro2018dynamical, martins2022universality, da2018topology}.

\section*{Author contributions statement}
H.V.R., D.D.L, A.A.B.P., A.F.M., B.R.d.C., S.G., E.K.L, Q.S.H., and M.P. designed research, performed research, analyzed data, and wrote the paper.

\section*{Competing interests}
B.R.d.C and Q.S.H. are employees of private companies. Views discussed in the manuscript do not represent the views or positions of their respective companies. There are no patents, products in development, or marketed products to declare. 

\section*{Acknowledgements}
We acknowledge the support of the Coordena\c{c}\~ao de Aperfei\c{c}oamento de Pessoal de N\'ivel Superior (CAPES -- PROCAD-SPCF Grant 88881.516220/2020-01), the Conselho Nacional de Desenvolvimento Cient\'ifico e Tecnol\'ogico (CNPq -- Grant 303533/2021-8), and the Slovenian Research Agency (Grants J1-2457 and P1-0403).

\bibliography{references.bib}

\end{document}